\documentclass{nature}
\usepackage[colorlinks=true,citecolor=blue,linkcolor=magenta]{hyperref}

\usepackage[markup=nocolor, authormarkupposition=left]{changes} 
\usepackage{soul}
\usepackage[utf8]{inputenc}
\usepackage[english]{babel}
\usepackage{amsmath,amsfonts,amssymb}
\usepackage[T1]{fontenc}
\usepackage{url}
\usepackage{amsmath}
\usepackage{siunitx}
\usepackage[version=4]{mhchem}
\usepackage{bm}

\usepackage{amsfonts}
\usepackage{amssymb}
\usepackage[font=small,labelfont=bf]{caption}
\usepackage{graphicx}
\usepackage{changes}
\graphicspath{{./Figures/}}

\newcommand{\Fref}[1]{Figure~\ref{#1}}
\newcommand{\fref}[1]{Fig.~\ref{#1}}

\newcommand{\acite}[1]{\cite{#1}}

\usepackage{setspace}
\title{Large-scale photonic chip based pulse interleaver for low-noise microwave generation}

\author{Zheru Qiu$^{1,2}$, Neetesh Singh$^{3,***}$, Yang Liu$^{1,2}$, Xinru Ji$^{1,2}$,	Rui Ning Wang$^{1,2,4}$, Franz X. Kärtner$^{3,**}$, Tobias Kippenberg$^{1,2,*}$}

\begin{document}

\maketitle

\begin{affiliations}
	\item Swiss Federal Institute of Technology Lausanne (EPFL), CH-1015 Lausanne, Switzerland
    \item Center for Quantum Science and Engineering, EPFL, CH-1015 Lausanne, Switzerland
    \item Center for Free-Electron Laser Science, Deutsches Elektronen-Synchrotron, 22607 Hamburg, Germany
    \item Currently with Luxtelligence SA, CH-1015 Lausanne, Switzerland
    \item[] *:\href{mailto://tobias.kippenberg@epfl.ch}{tobias.kippenberg@epfl.ch}, **:\href{mailto://franz.kaertner@desy.de}{franz.kaertner@desy.de}, ***:\href{mailto://neetesh.singh@desy.de}{neetesh.singh@desy.de}
\end{affiliations}

\begin{abstract}
Microwaves generated by optical techniques have demonstrated unprecedentedly low noise and hold significance in various applications such as communication, radar, instrumentation, and metrology.
To date, the purest microwave signals \acite{kalubovilage2022x, fortier2011generation, xie2017photonic} are generated using optical frequency division with femtosecond mode-locked lasers.
However, many femtosecond laser combs have a radio frequency (RF) repetition rate in the hundreds of megahertz range, necessitating methods to translate the generated low-noise RF signal to the microwave domain. 
Benchtop pulse interleavers \acite{Haboucha11, jiang2011noise} can multiply the pulse repetition rate, avoid saturation of photodetectors \acite{fortier2013photonic}, and facilitate the generation of high-power low-noise microwave signals, which have to date only been demonstrated using optical fibers \acite{Haboucha11} or free space optics.
Here, we introduce a large-scale photonic integrated circuit-based interleaver, offering a size reduction and enhanced stability. 
The all-on-chip interleaver attains a 64-fold multiplication of the repetition rate, directly translated from 216~MHz to 14~GHz in microwave Ku-Band.
By overcoming photodetector saturation with the presence of high peak power at lower repetition rates, the generated microwave power was improved by 36~dB, with a phase noise floor reduced by more than 10 folds to -160~dBc/Hz on the 14~GHz carrier.
The device is based on a few-mode low-loss (8~dB/m) and high density \ce{Si3N4} photonic integrated circuit fabricated by the photonic Damascene process \acite{liu2021high}.
Six cascaded stages of Mach-Zehnder interferometers with optical delay lines up to 33~centimeters long are fully integrated into a compact footprint of 8.5~mm×1.7~mm.
The lithographically defined precision of the optical waveguide path length enables the scaling up of the interleaved frequency to millimeter-wave bands, which is challenging the fiber-based counterparts.
This interleaver has the potential to reduce the cost and footprint of mode-locked-laser-based microwave generation, allowing for field deployment in aerospace and communication applications.
\end{abstract}
%%%%%%%%%%%%%%%%%%%%%%%%%%%%%%%%%%%%%
\section{Introduction}

\begin{figure*}[htp]
  \centering
  \includegraphics[width=0.95\textwidth]{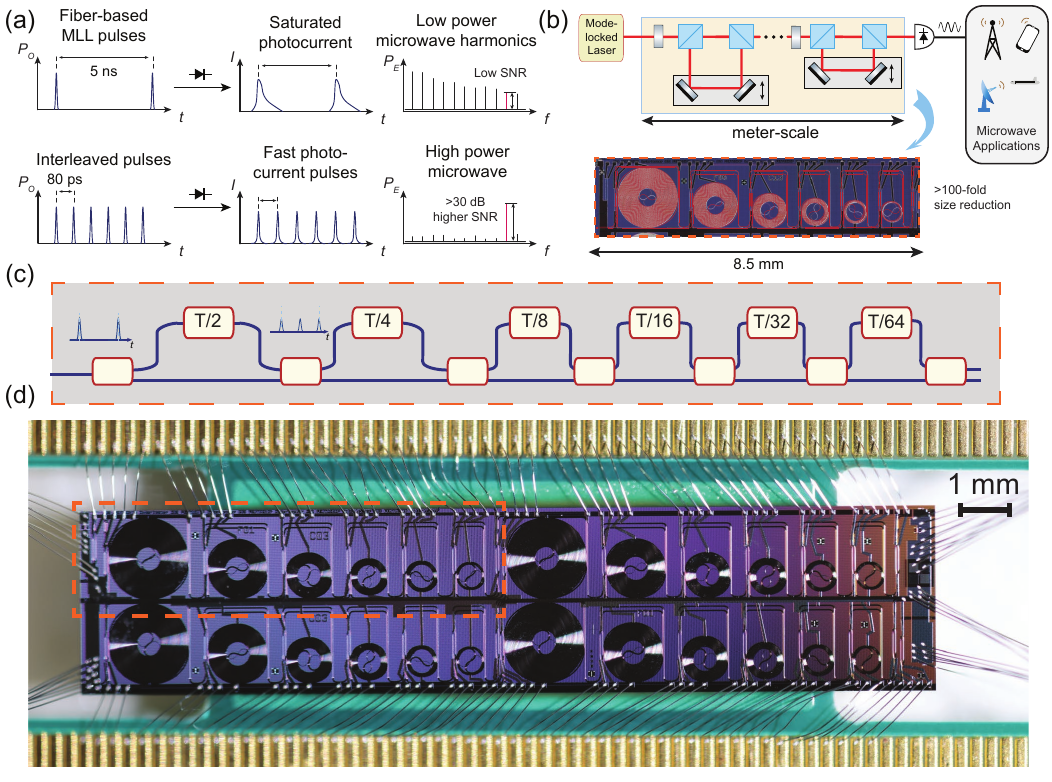}
  \caption{\footnotesize \textbf{Large-scale photonic integrated circuit based interleaver for low-noise microwave generation} 
    (a) Illustration of the photonic microwave generation process with a mode-locked fiber laser and the improvement of microwave power and SNR through pulse interleaving.
    (b) Size reduction of the pulse interleaver by replacing the free space optics with the on-chip components in a photonic microwave generation system.
    Inset is the optical micrograph of the fabricated \ce{Si3N4} interleaver chip, overlaid with the schematic of the \ce{Si3N4} waveguide.
    (c) Illustration of the working principle of the cascaded unbalanced Mach-Zehnder interferometer pulse interleaver.
    (d) Top view image of the 6-stage interleaver chip mounted on a printed circuit board for breaking out the heater connections.
}
\label{f1}
\end{figure*}

\begin{figure*}[ht]
  \centering
  \includegraphics[width=0.99\textwidth]{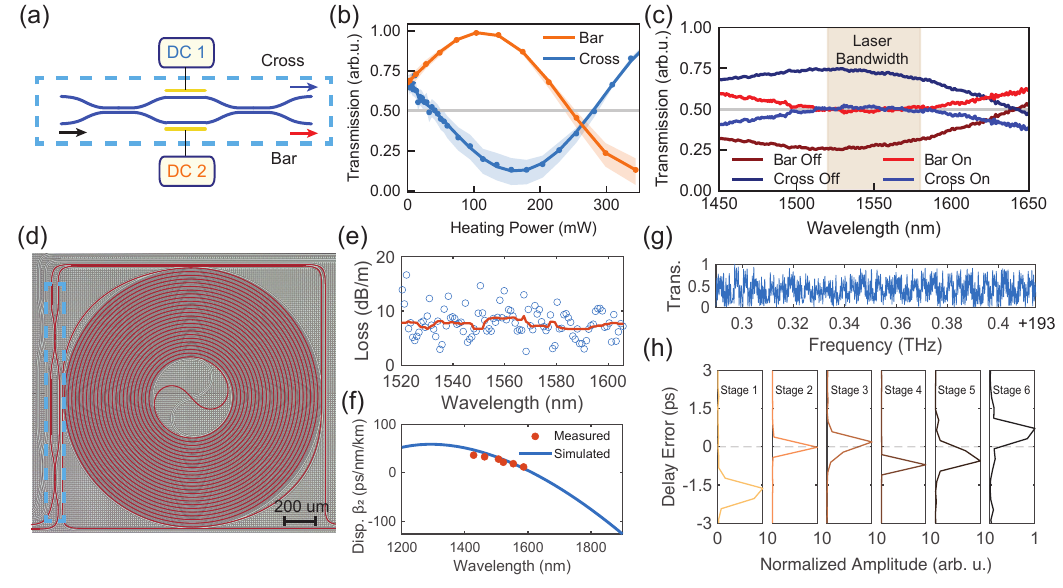}
  \caption{\footnotesize \textbf{Characterization of the \ce{Si3N4} photonic chip based pulse interleaver.}
  (a) Schematic of the balanced MZI tunable coupler implemented with two 50:50 waveguide directional couplers. DC: direct current.
  (b) Optical transmission through the bar port as a function of the heating power applied to either arm at 1550~nm.
  The shaded area shows the minimum to maximum range of the transmission within $1550\pm 10$~nm.
  (c) Normalized transmission spectrum of the tunable coupler when cold and having 187~mW heating applied to one arm.
  (d) Optical micrograph of a single interleaver stage, highlighting the tunable coupler in (a).
  (e) Waveguide propagation loss of the delay line section, characterized by the Lorentzian fit of ring resonator resonances, which gives the intrinsic loss rate.
  Circles represent individual resonances, and the red line is the moving median.
  (f) Simulated and characterized waveguide dispersion of the long delay line, described as $D=-2\pi c\beta_2/\lambda^2$, where $c$ is the speed of light and $\beta_2$ is the group velocity dispersion parameter.
  (g) Normalized transmission spectrum through the 6-stage interleaver near 193~THz, showing a part of the interference fringes used in the delay length characterization.
  (h) Time delay error characterized by the Fourier transformation of the optical transmission spectrum, zoomed in to the relevant frequency component indicating the time delay errors of each delay line.
  }
\label{f2}
\end{figure*}

High-purity microwave signals have a wide range of applications in communications \cite{koenig2013wireless, zou2021optoelectronic}, radar \cite{ghelfi2014fully, radarbook}, metrology\cite{santarelli1999quantum}, and analog-to-digital conversion\cite{adc_survey}. 
State-of-the-art low-noise microwave generation is based on compact dielectric resonant oscillators (DRO), which have been engineered for commercial purposes for decades.
In the laboratory environment, exceptionally low-noise microwave signals can be generated by optical means\cite{fortier2011generation, OFDScience}. In particular, low phase noise aspect of an ultra-stable optical reference can be transferred down to the microwave regime with a powerful technique known as optical frequency division (OFD), which has resulted in low-phase-noise microwaves reaching below -170~dBc/Hz at offsets $>10$~kHz \cite{fortier2011generation, xie2017photonic}.
In this technique, a solid-state or fiber-based mode-locked laser (MLL) is usually stabilized by locking a comb line to an ultra-low noise optical reference.
The stabilized repetition rate is then converted to a high-purity electronic signal by photodetection.

A challenge in practical applications of OFD is the gap between the pulse repetition rate and the microwave frequencies desired for utilization.
Most important applications such as communication systems, radar, and instrumentation require high-purity microwaves in the range of a few to tens of GHz.
At the same time, the repetition rate of the low-noise MLLs typically falls in the range of a few hundred MHz.
The short photocurrent pulses naturally contain high-order harmonics of the repetition rate, which can be used to bridge the gap in frequency.
However, the signal-to-noise ratio (SNR) of the generated microwave is limited by the photocurrent saturation in the photodiodes (PDs), even with state-of-the-art modified uni-traveling-carrier (MUTC) PDs \cite{fortier2013photonic}.
The PDs saturate with the incident of the high-peak-power femtosecond (fs)-pulses from MLLs, slowing down the response and thus limiting the achievable power of high-order harmonics.
While there is an elegant technique that bypasses this challenge by phase-locking to the rising edge of the photocurrent pulses\cite{hyun2020attosecond}, it necessitates careful control of laser intensity noises and high system complexity.
Soliton microcombs, on the other hand, offer a high-repetition-rate pulse train \cite{liu2020photonic}, but the currently achievable noise performance even with delicate feedback stabilization \cite{jin2024microresonator} is not yet comparable with conventional MLLs.

A common strategy to alleviate this problem is multiplying the optical pulse repetition rate using mode-filtering cavities \cite{diddams2009improved, jiang2011noise} or pulse interleavers\cite{Haboucha11}.
The pulse train with multiplied repetition rates has a lower peak power at the PD, which increases the generated microwave power by reducing saturation and improving the signal-to-noise ratio limited phase noise floor.
These pulse rate multipliers are used in the experiments demonstrating the generation of the purest microwave signal\cite{xie2017photonic} and commercial optical microwave generation systems\cite{8752385}.
However, the mode filtering cavities will reduce the available optical power and require servo locking.
The fiber-based and free-space interleavers have a large footprint  (\fref{f1} (b)), are prone to mechanical noise and temperature fluctuations, and necessitate careful adjustment for achieving desired delay lengths\cite{Matlis:23}.
%Moreover, the fiber interleavers can hardly scale to higher repetition rates $>100$~GHz, as the $<2$~mm delay lengths or length differences are difficult to precisely implement.

In this work, we demonstrate a miniaturized 6-stage pulse interleaver on a low-loss silicon nitride photonic integrated circuit (\fref{f1} (c)) fabricated at wafer scale.
By incorporating up to 33~cm long low-loss delay lines, this chip can multiply the repetition rate $f_{\mathrm{rep}}=\SI{216.7}{\mega\hertz}$ of a fiber-based 1550~nm MLL by $2^6$ times.
The device footprint of 8.5~mm$\times$1.7~mm device  (\fref{f1} (d)) is significantly smaller than fiber-based or free-space counterparts and is more than one order of magnitude smaller than the reported 4-stage low-index silica waveguide-based mode-filter \cite{sander201210}. %$<1/25$ of the size of
This device allows for low-cost mass production of compact ultra-low-phase-noise photonic microwave sources well suited for fiber-based MLLs or the future integration with chip-based MLLs \cite{singh2020towards, singh2023chip}.

\section{Device design and fabrication}

\begin{figure*}[htbp]
  \centering
  \includegraphics[width=0.95\textwidth]{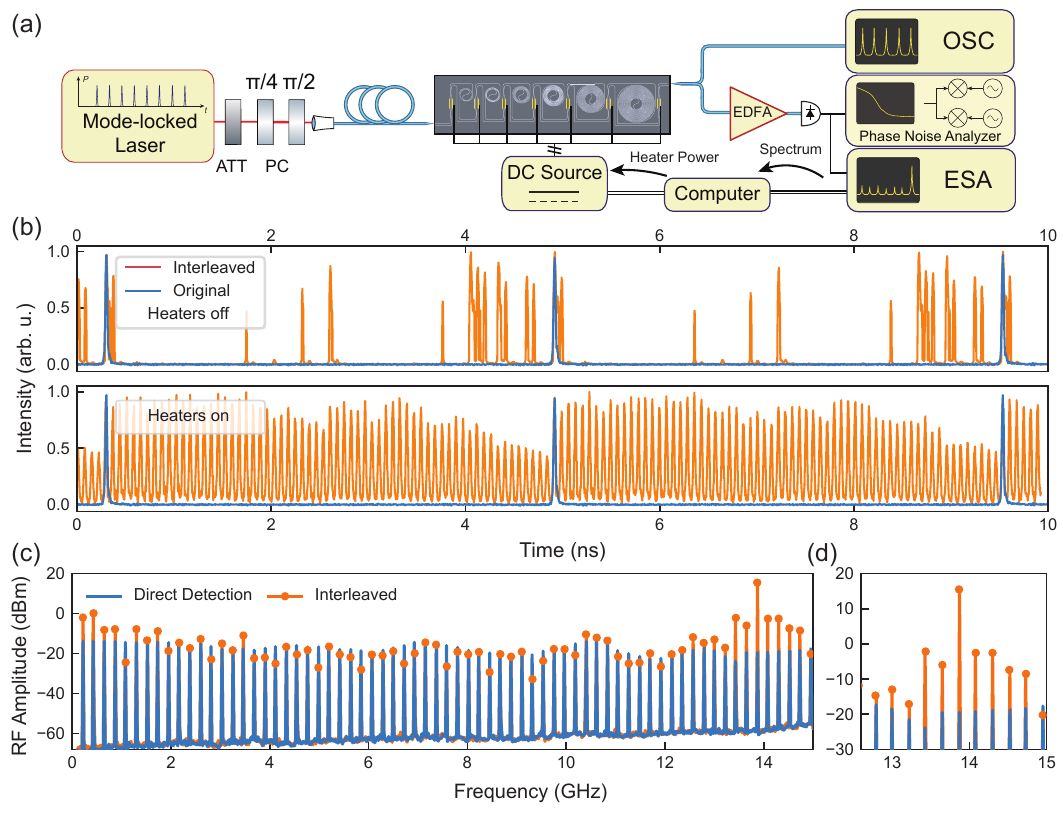}
  \caption{\footnotesize \textbf{Demonstration of 6-stage pulse interleaving using a femto-second mode locked laser.}
  (a) The experiment setup for the pulse interleaving demonstration, where the 217~MHz pulse from the erbium doped fiber based mode-locked laser is interleaved.
  ATT: attenuator, PC: polarization controller, ESA: radio frequency (RF) spectrum analyzer.
  (b) Time domain waveform of the pulses before and after interleaving, with the optimized heating power of the tunable couplers not applied (lower) and applied (upper).
  (c) Spectrum of the microwave signal generated by photodetection of the pulse train with and without interleaving, where the average optical power is set at 76~mW and 1.7~mW, respectively, for the maximum 13.9~GHz power in both cases.
  (d) Close-in view of the spectrum near the desired 13.9~GHz microwave, highlighting the 36~dB increase in the produced power.
  }
\label{f3}
\end{figure*}

\begin{figure*}[htbp]
  \centering
  \includegraphics[width=0.95\textwidth]{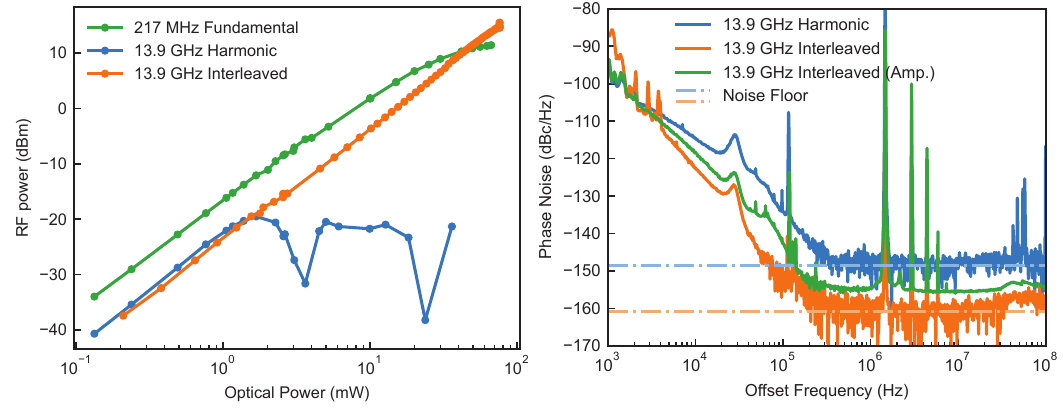}
  \caption{\footnotesize \textbf{Low-noise microwave generation using a large-scale photonic chip scale interleaver.}
  (a) The power of generated 13.9~GHz microwave as function of the average optical power after interleaving, which is shown together with the power of the 216~MHz fundamental radio frequency signal and the harmonic 13.87~GHz microwave generated by the pulse train without interleaving.
  This shows the saturation of the photodiode by the original pulse train and the higher power from the interleaved pulse train.
  (b) Phase noise of the generated 13.87~GHz microwave from the non-interleaved pulses, the interleaved pulses and the interleaved pulses amplified by an EDFA.
  The total shot and thermal noise floors are shown as dashed lines.
  }
\label{f4}
\end{figure*}

We adopt a cascaded unbalanced Mach-Zehnder interferometer (MZI) design (\fref{f1} (a)) for the interleaver.
In the $n$th stage of an ideal interleaver, each input pulse is equally divided into two halves by a coupler.
The pulse train in the longer MZI arm is delayed by $1/(2^nf_{\mathrm{rep}})$ and is subsequently combined with the pulse in the other half at the output coupler of each stage.
In the frequency domain, this pulse interleaving process modifies the relative phase of the comb lines of the MLL pulse train, while the amplitudes remain unchanged.
Therefore, for an ideal interleaver, there is no energy loss when summing up the pulse energies at both outputs.
The two outputs can be simultaneously utilized by separate photodetection and coherently combining the generated microwave.
This is in stark difference from the approaches that filter out modes \cite{jiang2011noise, sander201210} to increase the frequency interval, where the total energy of the pulse train is reduced to $1/m$ for a $m$-times repetition rate multiplication.

%In the Damascene fabrication of the \ce{Si3N4} photonics integrated circuits, the waveguide thickness and width may vary due to non-uniformities in the chemical ....
To compensate for the fabrication uncertainty in the coupler splitting ratio and unbalanced delay line losses for each interleaver stage, we employ balanced MZI tunable couplers composed of two 50:50 directional couplers and two arms integrated with \ce{Ti}/\ce{Pt} micro-heaters (\fref{f2} (a)).
The adjustment of the splitting ratio is achieved by applying different phase shifts to the two arms (\fref{f2} (b)).
\Fref{f2} (c) shows the normalized transmission spectrum through the fabricated MZI directional couplers, tested with an in-house comb-assisted laser-sweeping-based optical network analyzer \cite{riemensberger2022photonic}.
The transmission can be tuned to $50\pm4\%$ within the $1550\pm 30$~nm range, ensuring equal splitting of all the spectral components of the MLL comb.
%\hl{(add the quantified discription of the termal tuning efficiency.)}
We anticipate that the use of thermal isolation trenches for future development can enhance tuning efficiency \cite{yong2022power}.
The \ce{Si3N4} photonic integrated circuits are fabricated with the photonic damascene process \cite{liu2021high}.
The delay line waveguides are designed to have a $\SI{1.5}{\micro\meter}\times\SI{700}{\nano\meter}$ cross-section which is narrower than previous works to reduce the coupling to high-order modes; as a trade-off, the waveguide scattering loss is higher than wider waveguides.
%\hl{(do you have the SEM of the waveguide cross section, it can be a figure inset)}
The total waveguide propagation loss is in the range of 8 to $10~\mathrm{dB/m}$ (\fref{f2}).
This low waveguide loss only contributes to a total on-chip optical loss of 2.2~dB, allowing for minimal degradation in SNR.
\Fref{f2} (f) shows the simulated and measured (Supplementary Note 1) waveguide dispersion, respectively.
The signal wavelength range lies in the weak anomalous dispersion regime.

The lithography can define precise optical delay length for the interleaver, which is critical to ensuring low errors of time delay that can otherwise corrupt the regular timing between the pulses and negate the phase noise reduction by shot noise correlations of the short pulses \cite{quinlan2013exploiting}.
On the interleaver chip, the time delay is determined by the waveguide physical length and group index, which are precisely defined by deep ultraviolet lithography.
We characterized the time delay of each interleaver stage on fabricated chips by performing a Fourier transform of the transmission spectrum of the interleaver (Supplementary Note 2).
\Fref{f2} (h) shows the longest delay time of 2308~ps (corresponding to 33~cm) with a timing error of $<2$~ps (0.12\%).
%\hl{(explain why stage 6 (short length) has a higher delay errors? it is more than 0.4 ps.)} _ good catch... It's a mistake in the labeling, should be inverted.
For shorter delay lengths, the delay time errors are smaller and close to the resolution limit of this characterization technique (0.4~ps, limited by the analysis wavelength range of 1540 to 1560~nm).
The $<2$~ps delay error is comparable to or even better than the hand-polished fiber delay lines, which can minimize microwave power loss and the increase in phase noise\cite{quinlan2014optical}.
We note that inserting more stages with shorter delay lengths of a few mm or hundreds of \SI{}{\micro\meter} is difficult with fiber delay lines but can be simply implemented on a chip.
In contrast, using photonic integrated circuits, the repetition rate of the generated pulse train can be easily extended to $>100$~GHz \cite{MMWMLLInterleaver}, enabling ultra-low phase noise millimeter wave generation with a simple setup.
The same design may also be implemented with 200~nm thick etched \ce{Si3N4} waveguides, which have a larger mode field diameter and can reduce nonlinear broadening.
The subtractive fabrication process may also yield better control of the waveguide dimension, further improving the delay precision.

\section{Pulse interleaving and low-noise microwave generation}

We demonstrate the pulse interleaving using \SI{216.7}{\mega\hertz} repetition rate pulses from a free-running erbium-doped fiber-based MLL (OneFive Origami).
After on-chip interleaving, the pulse train was directed to a MUTC PD (Freedom Photonics FP1015A, 5.3~V bias) for microwave generation.
We optimized the heating power of the 7 tunable couplers to maximize the suppression of undesired harmonics on the RF spectrum (see Methods).
\Fref{f3} (b) illustrates the generation of 64 regular time-delayed copies of input pulses with optimal phase shift settings.
The undesired harmonics on the RF spectrum could be suppressed by more than 14~dB (\fref{f3} (c)) relative to the generated microwave level.
The suppression ratio was limited by optical loss in long delay lines, which caused parasitic amplitude modulation of the pulse train (as seen in \fref{f3} (b)).
%Need to double check: \added{due to the fact the pulse delayed in the largest interleaver arm experienced the highest loss.}
This effect results in sidebands at $\pm433$~MHz offset in the output microwave.
We should note that this amplitude modulation doesn't degrade signal phase coherency and has minimal impact on applications.

\Fref{f4} (a) shows the RF power as a function of the average optical power at the PD.
The non-interleaved pulses saturated the PD at $<\SI{1}{\milli\watt}$ average optical power and limited the generated 13.87~GHz harmonic power to $<-19~\mathrm{dBm}$.
The interleaved pulses could generate more than $16~\mathrm{dBm}$ microwave power with 76~mW of average optical power when an erbium-doped fiber amplifier (EDFA) was used to amplify the signal before photodetection.
This corresponds to $>36~\mathrm{dB}$ improvement in generated microwave power, only limited by the power handling of the photodetector.
When a lower-cost PIN photodetector was used instead of the MUTC diode,  $>10~\mathrm{dB}$ improvement in power was also achieved (Supplementary Note 5).
The microwave phase noise was characterized by a phase noise analyzer (R\&S FSWP) with a thermal noise limited noise sensitivity after 1000 times of cross-correlation.
As shown in \fref{f4} (b), with the interleaver we reach the best phase noise performance of -160~dBc/Hz at 1~MHz offset for the generated -15~dBm 13.87~GHz microwave signal at ${\sim1.2~\mathrm{mA}}$ photocurrent.
This noise floor is limited by the continuous wave shot noise and thermal noise floor \cite{jiang2011noise} shown by the dashed line in \fref{f4} (b), and is ${\sim 12~\mathrm{dB}}$ lower than the noise floor reached at the same photocurrent without using pulse interleaving.
Here, the EDFA was not used to avoid extra noise introduced by pulse broadening induced by the Kerr nonlinearity in waveguides with tight optical mode confinement and spontaneous emission of the EDFA.
When the output pulse train was amplified by the EDFA, the phase noise floor was slightly raised to -155~dBc/Hz at $16~\mathrm{dBm}$ microwave power.
% caused by noise and pulse broadening caused by the dispersion in the EDFA.
At offset frequencies between $5~\mathrm{kHz}$ and $100~\mathrm{kHz}$, we also observed reduced phase noise in the interleaved pulse train.
This phase noise advantage is likely due to the decreased phase noise converted from amplitude noise \cite{fortier2013photonic, davila2018optimizing}, as the photodiode was less saturated by the interleaved pulses with lower peak power.
The slight increase of phase noise at the low offset frequencies less than $5~\mathrm{kHz}$ may be attributed to the low-frequency environmental perturbations on delay lengths and fiber-to-chip coupling which can be mitigated by packaging the interleaver chip and using low-noise DC sources.

\section{Summary}

We present a large-scale integrated pulse interleaver on a \ce{Si3N4} photonic integrated circuit platform.
With low-loss waveguide delay lines, this 8.5~mm$\times$1.7~mm device can efficiently multiply the pulse repetition rate from 217~MHz to 14~GHz on a single chip. Applying this integrated photonic multiplier to a free-running femtosecond laser enabled an increase in the generated microwave power by 37~dB and a reduction of the phase noise floor by 12~dB.
This integrated interleaver holds potential for future miniaturized ultra-low-phase-noise photonic microwave generators with fiber-based and can be co-integrated with chip-based mode-locked lasers.

\section*{Methods}
\begin{footnotesize}

\noindent\textbf{Details on Sample Fabrication}:
The photonics integrated circuits were fabricated with the photonic Damascene process with \ce{SiO2} and \ce{SiNx} hardmask for preform etching\cite{liu2021high}.
The waveguide width was reduced by \SI{150}{\nano\meter} on the mask data to compensate for an offset introduced in the \ce{SiNx} hardmask etching.
For the microheaters, a \SI{25}{\nano\meter} \ce{Ti} adhesion layer and \SI{500}{\nano\meter} of \ce{Pt} were sputtered on top of the photonics integrated circuit wafers with \SI{3}{\micro\meter} of \ce{SiO2} cladding.
Direct-write ultraviolet lithography with \SI{3}{\micro\meter} AZ ECI3027 photoresist was applied to define the \SI{5}{\micro\meter} wide heater lines.
We empirically find that the heater lines are more robust at high power when fabricated to be wide and thick.
The heaters were etched with an \ce{Ar} ion-beam etcher (Veeco Nexus IBE350) at $-30^\circ$ substrate angle.

\noindent\textbf{Details of the experimental setup for pulse interleaving}:

The interleaver can operate with the pulses entering either from the 6th stage or from the 1st stage.
We sent the pulse to the 6th stage in the experiments to reduce the excessive nonlinear broadening of the pulses in the \ce{Si3N4} waveguides.
In this way, the pulses were split before entering the longer delay lines, reducing the peak power and further mitigating nonlinear broadening (Supplementary Note 3).
A 20~cm single mode fiber delay line was used for pre-chirping the pulse (specified for 172~fs pulse duration) before coupling onto the interleaver chip.
The time domain characterization of the resulting pulse was performed using a sampling oscilloscope with a 34~GHz optical input (Keysight 86105D), triggered by the synchronization output of the mode-locked laser.

\noindent\textbf{Automatic optimization of the splitting ratios}:

The power splitting ratios of the 7 MZI tunable couplers between the delay lines (\Fref{f2} (a) and (b)) were controlled by the heating power $\left|p_n\right|$ applied to one of the two arms, the sign of $p_n$ decides which arm is heated.
The 14 heaters were driven by a multichannel DC source measure unit (nicslab).
In order to find the optimal set points of all 14 heaters for microwave generation, where the power of the 64th harmonic is maximized while the other harmonics are suppressed, we monitored the RF spectrum of the photodetected pulses with an ESA (Keysight N9030A).
We define a merit function $g=P_{64}-\sum_{n=1}^{63}P_n$, where $P_n$ is the power of the $\SI{n\times 217}{\mega\hertz}$ harmonic in dBm unit, measured by the ESA.
We employ the Nelder–Mead algorithm \cite{NM}, as implemented in \verb|scipy|, to optimize for the maximum $g$ as a function of $p_n$.
The Nelder–Mead algorithm, being a derivative-free simplex optimization method, is well-suited for optimizing our experimentally measured, noisy target function.
In our experiment setup with non-polarization-maintaining fiber, the polarization of the beam coupled on chip may not be perfectly aligned with the chip plane (for TE mode coupling).
The misalignment also affects the extinction ratio and the merit function.
For this, the initial automatic optimization was performed iteratively with manual fine tuning of the fiber polarization controller.
We characterized the on-chip optical loss after the optimization of polarization and the application of the optimal heater setting.
The 4.7~dB fiber-to-chip coupling loss and the 3~dB loss caused by the equal splitting of the output power from the two ports are subtracted.

\noindent \textbf{Funding Information}:
This work was supported by the EU H2020 research and innovation program under grant No. 965124 (FEMTOCHIP). 

\noindent \textbf{Acknowledgments}: 
The samples were fabricated in the EPFL center of MicroNanoTechnology (CMi).

\noindent \textbf{Author contributions}: 
Z.Q. and N.S. performed the experiments.
Z.Q., N.S., and Y.L. carried out data analysis and simulations. 
Y.L. and X.J. provided experimental supports.
Z.Q., Y.L and N.S. designed \ce{Si3N4} photonic integrated circuits.
R.N.W., Z.Q. and X.J. fabricated the \ce{Si3N4} samples.
Z.Q. wrote the manuscript with the assistance from N.S. and Y.L., with the input from all co-authors. 
F.X.K. and T.J.K supervised the project.

\noindent \textbf{Data Availability Statement}: 
The code and data used to produce the plots within this work will be released on the repository \texttt{Zenodo} upon publication of this preprint.

\noindent \textbf{Competing interests}
T.J.K. is a cofounder and shareholder of LiGenTec SA, a start-up company offering \ce{Si3N4} photonic integrated circuits.

\end{footnotesize}
%\vspace{-0.3cm}

%% NOTE: RevTeX only compiles with BibTeX, Configure your system to use it ONLY!

\pretolerance=0
\bigskip
\bibliographystyle{naturemag}
\bibliography{library}
\end{document}

\title{Supplementary Information for: \\ Large-scale photonic chip based pulse interleaver for low-noise microwave generation}

%\author{Zheru Qiu$^{1,2}$, Neetesh Singh$^{3}$, Yang Liu$^{1,2}$, %
%Xinru Ji$^{1,2}$, Rui Ning Wang$^{1,2,4}$, Franz X. Kärtner$^{3,\ast\ast}$, Tobias Kippenberg$^{1,2,\ast}$}

\author{Zheru Qiu$^{1,2}$, Neetesh Singh$^{3,***}$, Yang Liu$^{1,2}$, Xinru Ji$^{1,2}$,\\
Rui Ning Wang$^{1,2,4}$, Franz X. Kärtner$^{3,**}$, Tobias Kippenberg$^{1,2,*}$}

\address{${}^1$Swiss Federal Institute of Technology Lausanne (EPFL), CH-1015 Lausanne, Switzerland\\
${}^2$Center for Quantum Science and Engineering, EPFL, CH-1015 Lausanne, Switzerland\\
${}^3$ Center for Free-Electron Laser Science, Deutsches Elektronen-Synchrotron, 22607 Hamburg, Germany\\
${}^4$ Currently with Luxtelligence SA, CH-1015 Lausanne, Switzerland%
}
%\begin{abstract}
%\textbf{.}
%\end{abstract}

%%%%%% RESET EQUATION NUMBERS ETC. %%%%%%%%
\setcounter{equation}{0}
\setcounter{figure}{0}
\setcounter{table}{0}

\setcounter{subsection}{0}
\setcounter{section}{0}

% \begin{abstract}
% Supplementary Information accompanying the manuscript containing performance comparison with prior works, sample fabrication process,  simulations of ion doping, theoretical analysis of optical gain and characteristic parameters, experimental details including measurement methods, calibrations and  characterizations.
% \end{abstract}

\maketitle
\vspace{-4em}
{\hypersetup{linkcolor=black}\tableofcontents}
%\newpage

%%%%%%%%%%%%%%%%%%%%%%%%%%%%%%%%%%%%%%%%%%%%%%%%%%%%%%%%%%%%%%%

\newcommand\blfootnote[1]{%
  \begingroup
  \renewcommand\thefootnote{}\footnote{#1}%
  \addtocounter{footnote}{-1}%
  \endgroup
}

\blfootnote{%
* \href{mailto://tobias.kippenberg@epfl.ch}{tobias.kippenberg@epfl.ch}\\
** \href{mailto://franz.kaertner@desy.de}{franz.kaertner@desy.de}\\
*** \href{mailto://neetesh.singh@desy.de}{neetesh.singh@desy.de}%
}

\section{Dispersion of the \ce{Si3N4} waveguides}

\begin{figure}[htb]
	\centering
	\includegraphics[width=0.8\textwidth]{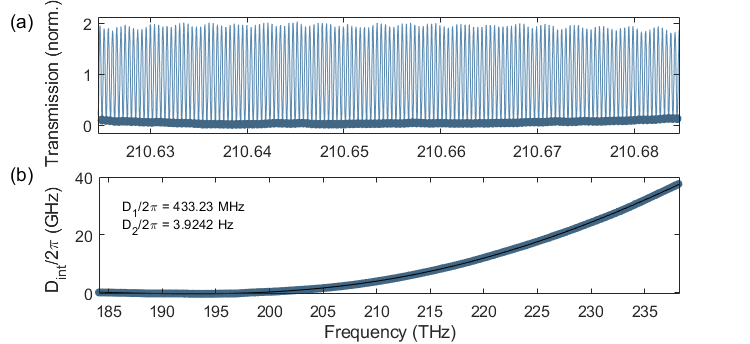}
    % D10602F03C06-1.1.1-1.1.1_pol1_laser1_calibData_disp_mod
	\caption{\textbf{Dispersion characterization} (a) High resolution transmission spectrum of the single stage MZI test device with the local minima labeled.
    (b) The fitted integrated dispersion curve of the points in (a).}
	\label{Fig:Disp}
\end{figure}

We characterized the dispersion of the delay lines by analysing the frequency-comb calibrated transmission spectrum of a single stage Mach-Zehnder interferometer interleaver (device number \verb|D10602F03C06|).
We located the local minima in the transmission spectrum and analysed the difference of their positions from a regular grid with a repetition rate of $D_1/(2\pi)$, which were then represented as the integrated dispersion $D_\mathrm{int}$ as shown in \Fref{Fig:Disp}.
\begin{align}
    D_1 & = \frac{1}{2}(\omega_{+1}-\omega_{-1})=\Delta\omega_{FSR}=\frac{2\pi c}{n_gL_R}=\frac{2\pi}{\beta_1L_R} = 2\pi \times 433.23\,\mathrm{MHz}\\
    D_2 & = \omega_{+1}-2\omega_0+\omega_{-1}=-\frac{4\pi^2\beta_2}{\beta_1^3L_R^2}=-\frac{\beta_2D_1^2}{\beta_1} =  2\pi \times 3.9242\,\mathrm{Hz}
\end{align}
From the polynomial expansion coefficients $D_i$ of the $D_\mathrm{int}(\mu)$ as a function of the number $\mu$ of the local minima, we extracted the group index of the waveguide to be $n_g=2.0977$ and the group velocity dispersion constant $D$ to be $+18.3\,\mathrm{ps/(nm\cdot km)}$ (anomalous).
The $D$ is close to the $\sim+18.0\,\mathrm{ps/(nm\cdot km)}$ specified value of standard SMF-28 telecom fiber, which accumulates to 0.3~ps for the maximum propagation length and 30~nm bandwidth.
% outer radius = 816.64u, GDS length difference = 329878.88 um, fitted frep = 433.24 MHz -> ng = 299792458/329878.88e-6/433.24e6 = 2.09767
% $D_1=\frac{1}{2}(\omega_{+1}-\omega_{-1})=\Delta\omega_{FSR}=\frac{2\pi c}{n_gL_R}=\frac{2\pi}{\beta_1L_R}$
% $D_2=\omega_{+1}-2\omega_0+\omega_{-1}=-\frac{4\pi^2\beta_2}{\beta_1^3L_R^2}=-\frac{\beta_2D_1^2}{\beta_1}$
% D1=433.24e6*(2*np.pi); D2=3.6553*(2*np.pi)
% -> beta1 = 1/329878.88e-6/433.24e6= 6.997081644604117e-09
% -> beta2 = beta'' = -D2*beta1/(D1**2) = -2.1687176062757758e-26 (SI, s^2/m/rad, 1e25 ps^2/km) 
% D -> -(2*np.pi*299792458)/(1550e-9**2)*beta2 = 1.7003572604002787e-05 (SI) = +17.00 ps/(nm⋅km) (= 1e-06) (anomalous GVD)
% 17*30*0.33e-3*(1+1/2+1/4+1/8+1/16+1/32) = 0.3313 (ps)
The simulated dispersion parameters in the main text were obtained with a mode solver (Lumerical) using the Sellmeier fit on the experimentally measured refractive index of the \ce{Si3N4} film (Woollam).

\section{Characterization of the time delays in the interleaver}
We take advantage of the frequency domain response of the cascaded MZI interleaver to precisely determine the delay lengths.
We perform high-resolution optical spectroscopy with a home-built optical vector network analyzer \cite{riemensberger2022photonic}.
We can prove that the delay length of the longest stage in the cascaded MZI device can be characterized by the Fourier transform of the optical transmission spectrum, where the time delay is represented as the position of the highest frequency peak in the transformed domain.

The complex transfer function of a cascaded MZI interleaver with $N$ stages can be modeled by the transfer matrix formalism:

\begin{equation}
    \mathcal{C}_n = \begin{pmatrix}
            \sqrt{1-k_n} & i\sqrt{k_n}\\
            i\sqrt{k_n}& \sqrt{1-k_n}
           \end{pmatrix},
    \mathcal{P}_n = \begin{pmatrix}
            \mathrm{e}^{i\beta(f) L_n} & 0\\
            0& 1
           \end{pmatrix}
\end{equation}
\begin{equation}
    \mathcal{T} = \prod^N_{n=1} \mathcal{C}_n\mathcal{P}_n\mathcal{C}_7
\end{equation}
\begin{equation}\label{eq1}
    \mathcal{T}_{11} = \sum_{k=0}^{2^N-1} a_k \exp\left(i\beta(f) \cdot\sum_n L_n s_{kn}\right)
\end{equation}

,where the $\mathcal{C}_n$ is the transfer matrix of the tunable directional coupler with a coupling ratio of $k$, and the $\mathcal{P}_n$ is the phase shift experienced in the nth stage delay line of propagation constant $\beta(f) =nf/c$ and length $L_n$.
Here we neglect the effect of dispersion in the analysis window.
The transfer matrix $\mathcal{T}_n$ of the entire 4 port system is the product of the transfer matrices corresponding to each stage.
The complex field transmission of a single port $\mathcal{T}_{11}$ to another can be written as \eqref{eq1} by collecting the terms with the same complex exponential, where $a_k$ are real constants and the $s_{kn}$ is the nth bit of the binary representation of integer $k$.

The transmitted intensity $T(f)$ is

\begin{align}
    T(f) &= \mathcal{T}_{11}\mathcal{T}^*_{11}\\
         &= \sum_k\sum_{k'} a_k a_{k'}\exp\left(i\beta(f)\cdot\left(\sum_n^N L_n s_{kn}-\sum_n^N L_n s_{k'n}\right)\right)
\end{align}

Let $L_n = 2^{n-1}L_1+\Delta L_n$, where the $\Delta L_n = c\Delta t_n/n_g$ is the error in the delay length.
We can see the terms in $T$ with the factor of $\exp(i\beta L_1\cdot2^{N-1})$ can only come from the $n=N$ case in the summation, as $\sum_{n=1}^{N-1} 2^{n-1} = 2^{N-1}-1<2^{N-1}$, such that these terms will come together with the factor of $\exp(i\beta \Delta L_N)$.
When performing a Fourier transform to $T(f)$, the highest "frequency" peak would be at $n_g(2^{N-1} L_1+\Delta L)/c=2^{N-1}n_gL_1/c+\Delta t_N$.
The $\Delta t_N$ is then derived as the offset between the peak in the "frequency" domain (or "delay time domain") spectrum and the demanded delay length by the mode-locked laser used in the experiment.

We extracted the delay lengths of the shorter stages by repeating the same characterization after disabling the longer stages with focused ion beam (FIB).
The FIB cuts were performed on the spirals with an Xe ion beam (FEI Helios G4).
The chip used in the experiment was from the same wafer of the characterized chip  (Number \verb|D10701|, Field 1 and 4).
Multiple waveguides were simultaneously cut at an angle, ensuring a complete loss of transmission and a low back reflection.

% Measured the one with spiral outer radius of [816.55950314, 601.71411887, 458.16530089, 366.11537501, 310.1606287 , 278.09599884], #+100
% Designed delay length is (offset = 100)
%# target_diff = np.r_[32.9769, 16.4884, 8.2442, 4.1221, 2.06105, 1.0305]*1e4 # in um
%# target_diff += offset*np.array([1/2**(n+1) for n in range(len(target_diff))])
% = [329819. , 164909.    ,  82454.5   ,  41227.25  ,  20613.625 ,        10306.5625]
% delay length is designed with n_g = 2.10 and actual fitted value is 2.09767 (-0.11% off)
% the f0 used in evaluation = 216.685e6 (f_rep of the Origami MLL)
% designed FSR = 299792458/329819e-6/2.1/2 = 216.419e6 -> error of 0.12% from MLL freq, 5.672 ps ....

\section{Optical spectrum of input and interleaved pulse trains}

\begin{figure}[htb]
	\centering
	\includegraphics[width=0.6\textwidth]{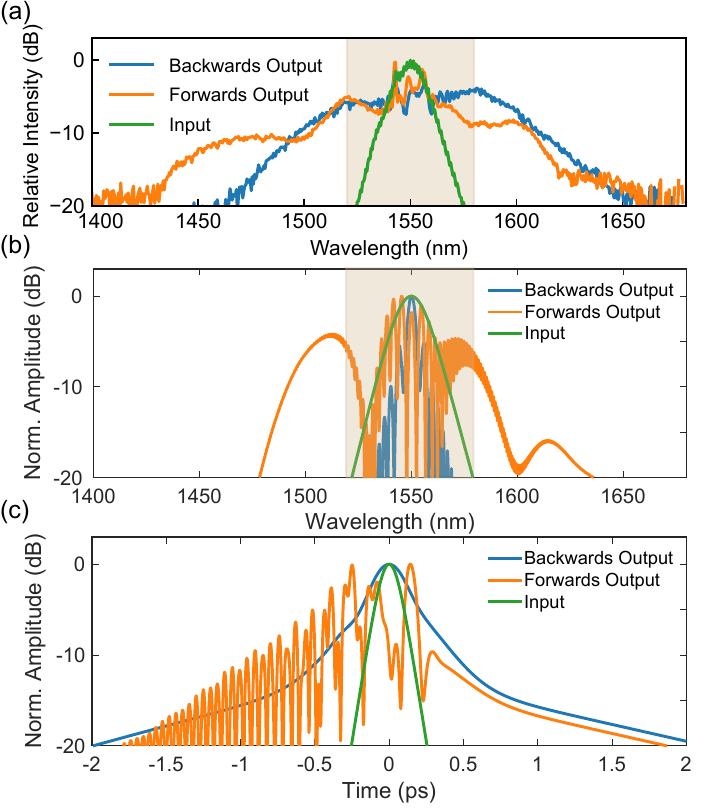}
	\caption{\textbf{Nonlinear Spectrum Broadening}
    (a) Measured optical spectrum of the input pulse train and the output from the interleaver chip in backwards and forwards operation.
    (b) Output optical spectrum simulated by numerical solution of the nonlinear Schrödinger equation (NLSE).
    (c) The simulated output pulse shape in time domain.
    %add input for comparsion *neetesh
    }
\label{Fig:SpectrumBroadening}
\end{figure}

\begin{figure}[htb]
	\centering
	\includegraphics[width=0.6\textwidth]{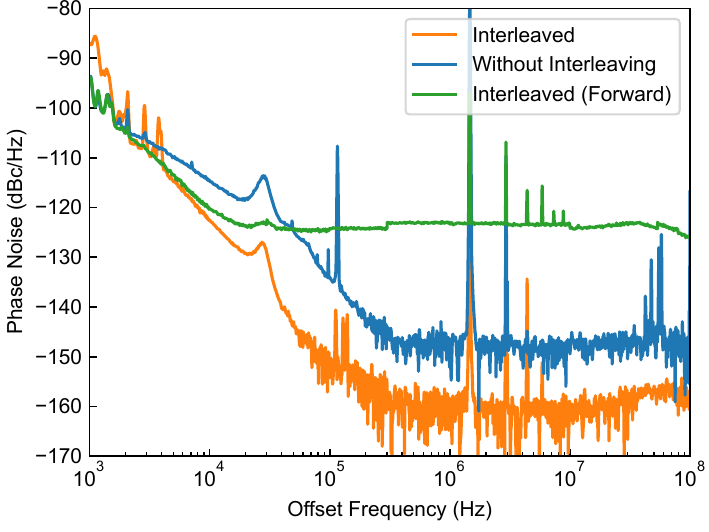}
	\caption{Measured phase noise of the generated 13.87~GHz microwave, comparing forwards operation and backwards operation of the interleaver.}
\label{Fig:PNFwd}
\end{figure}

\Fref{Fig:SpectrumBroadening}(a) shows the measured optical spectrum of the input pulse train from the mode-locked erbium fiber laser and the output from the interleaver chip.
We tested the device by launching the pulse from the forward and backward direction.
In the backward direction, the pulse goes through the short interleaver stages first. Whereas in the forward direction, the pulse starts from the long interleaver stages.
In the red and orange traces, the pulse input is at the shortest stage (backwards) and the longest stage (forwards) of the interleaver, respectively. 
The coupled total output power from the chip was kept at $\sim3.1$~mW.
In both cases, the laser spectrum is broadened by the nonlinear processes in the \ce{Si3N4} delay lines.
The broadening in the forwards case has distinctive sideband structures at $\sim 1430$~nm and $\sim 1600$~nm, which may indicate that there is a generation of incoherent supercontinuum.

Pulse propagation simulation by solving the NLSE in the forward and backward direction with the split-step Fourier method confirmed the spectrum broadening in the waveguides (\fref{Fig:SpectrumBroadening}(b)).
In the simulation, We used the waveguide dispersion parameters of $\beta_2 = -2.09\times10^2 \mathrm{ps^2/m}$, $\beta_3 = -5.07\times10^{-4} \mathrm{ps^3/m}$ and a fixed launched power for both directions.
\Fref{Fig:SpectrumBroadening}(c) shows that in the case of forwards operation, the pulses randomly break up in the time domain and disrupt the regular interval timing, while the pulses have retained the single Gaussian shape in backwards operation.
As we can see in the forward direction the bandwidth of the spectrum is broader and the temporal shape of the pulse is more structured than in the backward direction.
That is because of the stronger nonlinear effect experienced by the pulse in the forward direction compared to that in the backward direction.
The pulse enters with relatively higher power into the long sections of the interleaver from the forward direction compared to the backward direction, that is because before the pulse reaches the longer sections of the interleavers from the backward direction it has been split multiple times by the directional couplers present at every stage.
The higher power in the forward direction causes the pulse to experience higher nonlinear effects and thus shorter soliton fission length, as the fission length is inversely proportional to the peak power ($1/\sqrt{P}$) \cite{RevModPhys.78.1135, Singh:19}.

\Fref{Fig:PNFwd} shows the phase noise of generated microwave measured when the device is in forwards operation in comparison to the case of no interleaving and backwards operation.
In forwards operation, the noise floor degraded to $\sim-125$~dBc/Hz due to the incoherent nonlinear process.
In future work, to prevent modulation instability and pulse splitting, the waveguide GVD can be modified to normal by reducing the waveguide thickness or reducing the waveguide width.

\section{Average photocurrent saturation at high optical power}

\begin{figure}[htb]
	\centering
	\includegraphics[width=0.6\textwidth]{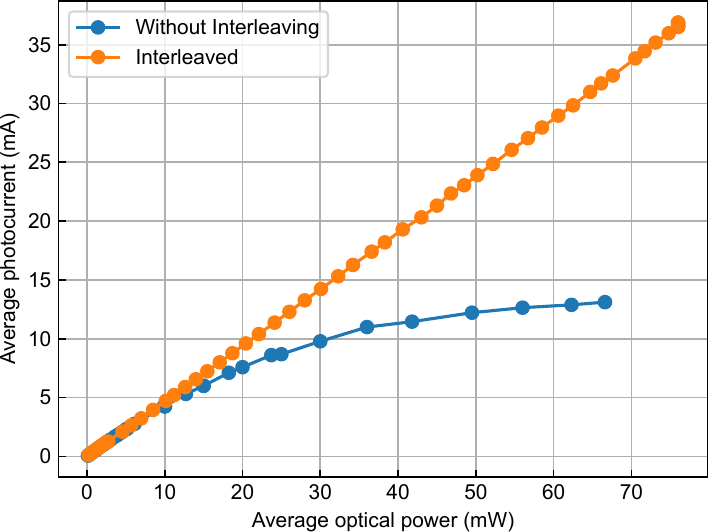}
	\caption{Average photocurrent as a function of average input power at the MUTC photodetector.}
	\label{Fig:Sat}
\end{figure}

\begin{figure}[htb]
	\centering
	\includegraphics[width=0.6\textwidth]{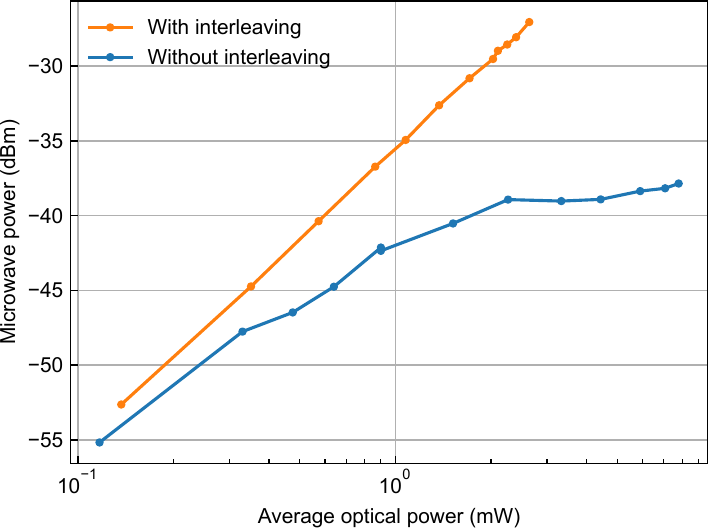}
	\caption{Generated 13.9~GHz microwave power with the Thorlabs DXM30AF photodiode as a function of the input average optical power.}
	\label{Fig:PINSat}
\end{figure}

We observed the average photocurrent also shows saturation at high input average power when the non-interleaved pulses are sent to the MUTC PD.
Meanwhile, the interleaved pulses were not causing appreciable saturation for $<75~\mathrm{mW}$ average power (\Fref{Fig:Sat}).
The average photocurrent was measured from the bias port of the Freedom Photonics FP1015A, at the same 5.3~V bias voltage as in the microwave generation experiment.

\section{Saturation behavior of a PIN photodetector}
PIN photodetectors are of lower cost than the MUTC photodetectors but are known to suffer more from saturation when illuminated with short pulses.
We observed a similar saturation of 64th harmonic power when the pulse is not interleaved when a Thorlabs DXM30AF photodetector is used for microwave generation.
As shown in \fref{Fig:PINSat}, the generated 13.9~GHz power saturated at $-37$~dBm.
For the interleaved pulse train from the chip, no power dependent saturation was observed up to $-27$~dBm generated power at 2.6~mW optical power.
The EDFA was not used in this experiment, which limited the average optical power and the microwave power in the interleaved case.
As no power dependent saturation was observed, we expect the advantage of interleaved pulses to be larger than 11~dB if the setup was optimized for higher output power or the pulses can be amplified.
We note that although the Thorlabs DXM30AF is specified with a higher responsivity (0.75~A/W at 1550~nm) than the Freedom photonics FP1015A MUTC PD ($0.6$~A/W at 1550 nm), the generated microwave power is 11~dB lower than the MUTC detector at 2.6~mW average power of interleaved pulses.
This indicates the PIN photodetector may have a stronger power-independent saturation when illuminated with short pulses.

%\clearpage
%%%%%%%%%%%%%%%%%%%%%%%%%%%%%%%%%%%%%%%%%%%%%%%%%%%%%%%%%%%%%%%%

%\section*{Supplementary References}
%\bigskip
% \renewcommand{\bibpreamble}{
% %$\dag$These authors contributed equally to this work.\\
% $\ast$\textcolor{magenta}{tobias.kippenberg@epfl.ch}
% }
\bibliographystyle{apsrev4-2}
\bibliography{library}